\documentclass[11pt]{article}

\usepackage[utf8]{inputenc}
\usepackage[T1]{fontenc}
\usepackage{lmodern}
\usepackage[margin=1in]{geometry}
\usepackage{graphicx}
\usepackage{float}
\graphicspath{{figures/}}
\usepackage{booktabs}
\usepackage{amsmath,amssymb}
\usepackage{authblk}
\usepackage{xcolor}
\usepackage{natbib}
\usepackage[hidelinks]{hyperref}

\newcommand{\rui}{\mbox{RUI}}
\newcommand{\ruimed}{\ensuremath{\rui_{\mathrm{median}}}}

\title{Quantifying research uniqueness with Unbalanced \\
Optimal Transport\\[3pt]
{\large Case studies of ten leading universities in neuroscience}}

\author[1]{Yu Ota\thanks{Corresponding authors: Yu Ota
(\texttt{s2510020@jaist.ac.jp}) and Amane Koizumi
(\texttt{akoizumi@jaist.ac.jp}).}}
\author[1,2]{Takayuki Ogi}
\author[1]{Yo Nakawake}
\author[3]{Daichi Mochihashi}
\author[1]{Amane Koizumi}
\affil[1]{Japan Advanced Institute of Science and Technology (JAIST),
Ishikawa, Japan}
\affil[2]{Fukushima University, Fukushima, Japan}
\affil[3]{The Institute of Statistical Mathematics, Tokyo, Japan}

\date{\today}

\begin{document}
\maketitle

\begin{abstract}
\noindent
Research evaluation has developed around quantity, citation impact and, more
recently, the accumulation of research excellence, but these dimensions remain
largely dependent on citations, which describe how research is received rather
than what it contains. We propose \emph{uniqueness} as a complementary dimension
and measure it with the Research Uniqueness Index (\rui{}), which asks how far a
university's research sits from what is typical in its field, using no citations
but the content of the research itself. Each researcher is represented by an
embedding of their abstracts, and an unbalanced version of Optimal Transport
returns the share of that work which cannot be matched into a random sample of
same-field researchers, the standard population; that unmatched share is the
score. Because every researcher is scored, a university is a distribution rather
than one number, and we treat that distribution as the indicator: it shows
whether uniqueness is spread across an institution or concentrated in a few
groups. Quantiles summarise it and, being quantiles, do not reward size. Since
uniqueness is defined only relative to a field, we demonstrate the index within
one, neuroscience (Scopus, 2021--2025; $9{,}861$ reference researchers). All ten
universities studied sit well above the standard ($\ruimed$ $0.40$--$0.53$ versus
$0.27$) and five specialised institutes higher still ($0.57$--$0.73$), the most
unique work often confined to a narrow niche that the same scores name. \rui{} is
uncorrelated with citation impact, and a controlled test confirms that it tracks
displacement from the field rather than topical concentration; results are robust
to an alternative embedding and replicate in computer science and economics. The
index is not intended as a league table: the gap above the standard and the
researcher-level niche structure are stable, whereas the order among adjacent
universities is not.
\end{abstract}

\noindent\textbf{Keywords:} research uniqueness; optimal transport; word
embeddings; neuroscience; research evaluation; institutional assessment;
responsible metrics; science mapping

\section{Introduction}
Scientific progress depends not only on the cumulative development of established
lines of inquiry but also on the capacity of researchers to pursue questions,
combinations and approaches that depart from the mainstream. Diversity in research
portfolios has long been regarded as important for maintaining the exploratory
capacity of science \citep{stirling2007}, and \citet{kuhn1977} described the
tension between tradition and innovation as essential to scientific development.
Consistent with this view, atypical combinations of prior knowledge have been
associated with unusual long-run scientific impact \citep{uzzi2013}.

Research evaluation, however, has developed largely around dimensions that are
only indirectly related to such intellectual distinctiveness. At the institutional
level, research performance has been described first in terms of quantity, such as
publication output, and then in terms of quality or impact, most commonly through
citation-based indicators. More recently, attention has turned to the depth of
research excellence within an institution: \citet{shirabe2021} proposed
substantiality as a construct for capturing the accumulation of research
excellence, rather than representing institutional performance by an average
alone. These developments have enriched the vocabulary of research evaluation, but
the dominant indicators of quality, impact and accumulated excellence remain
dependent on citations.

This dependence matters when the object of evaluation is the distinctiveness of the
research itself. Citations describe how research is subsequently received by a
scientific community; they do not describe the intellectual content of that
research, or the position it occupies within a field. Unconventional work may
moreover be recognised late or unevenly, which places novelty at a disadvantage in
citation-based evaluation \citep{wang2017}. Researchers may consequently be
encouraged towards safer and more conventional problems \citep{foster2015}, while
evaluation organised around established disciplinary categories can disadvantage
research that falls outside them \citep{rafols2012}. Research that departs most
strongly from the mainstream may therefore be precisely the research that
conventional indicators are least well equipped to characterise.

Existing measures of scientific novelty address part of this problem. Atypical
combinations of prior knowledge provide one influential approach
\citep{uzzi2013}, and later work has developed textual and semantic measures that
detect novelty directly from scientific documents
\citep{shibayama2021,zhao2025}. These measures are predominantly paper-level,
however. Many rely on citation or reference relations, and even
citation-independent semantic measures generally ask whether an individual paper
is novel, rather than whether a population of researchers occupies a distinctive
position relative to its disciplinary environment. What remains insufficiently
measured is how far an institutional research community, taken as a population,
departs from what is typical in its field.

We therefore propose \emph{uniqueness} as a complementary dimension of research
evaluation, conceptually distinct from research quantity, citation impact and the
accumulation of research excellence. By uniqueness we do not mean quality or
scientific merit, but the extent to which the research pursued by a researcher or
an institution occupies intellectual territory that is sparsely represented in the
surrounding field. Two questions follow: how far does a university's research
depart from what is typical in its discipline, and which researchers and research
areas account for that departure? Publication and citation counts answer neither,
because they measure volume and reception rather than intellectual position; nor
does within-unit diversity, which describes the spread of research inside a unit
rather than its displacement relative to an external disciplinary norm.

To address this gap we introduce the Research Uniqueness Index (\rui{}), a
citation-independent measure of the position of research content relative to a
same-field standard population. Each researcher is represented from the textual
content of their abstracts, and a university is treated as a population of such
researchers. We then use an unbalanced version of Optimal Transport
\citep{villani2009,peyre2019,chizat2018} to quantify the share of each
researcher's representation that cannot be matched into the standard population:
research located in densely populated regions of a field is readily matched,
whereas research occupying sparsely populated territory retains unmatched mass,
and that unmatched share is the researcher's score. Because every researcher is
scored, an institution is a distribution rather than a single number, and it is
that distribution we treat as the indicator; quantiles of it serve as summaries
which, being quantiles, do not reward size. Because uniqueness is defined only
relative to a field, the index must be computed within one; any field would serve,
and we use neuroscience as our demonstration.

Applied to ten leading universities, \rui{} places all ten above the standard
population, identifies the topics that set each apart---often a single, narrow
line of work---and places specialised institutes highest of all. The scores are
uncorrelated with citations and are not a restatement of topical diversity: a
controlled test shows \rui{} tracking a unit's displacement from its field rather
than the concentration of its topics, and the same patterns recur in computer
science and economics. The index is not intended as a league table. What is stable
is the gap above the standard, the broad grouping of institutions and the
researcher-level niche structure; the order among adjacent universities is
sensitive to resampling and to how each institution's boundary is drawn.

\section{Related work}
\label{sec:related}
\subsection{Measuring novelty, diversity, and uniqueness}
The Introduction placed uniqueness alongside quantity, impact and
substantiality as a dimension of research evaluation. Here we locate it among the
constructs used to describe research distinctiveness, which have been approached
mainly through novelty and diversity. \emph{Novelty} is
typically a paper-level property, an atypical combination of prior knowledge
\citep{uzzi2013}. In their review, \citet{zhao2025} classify novelty measures by
the data they use into three families: citation-relations-based (e.g.\ atypical
reference combinations, \citealp{uzzi2013,foster2015,wang2017}), textual or
semantic measures built from the words of a paper, and multi-type measures that
combine text with citations or networks \citep{shibayama2021}. \emph{Diversity}
and its inverse, specialisation or concentration, are within-unit properties, the
spread of a unit's work across sub-fields (Rao--Stirling diversity,
\citealp{stirling2007}; interdisciplinarity measures, \citealp{leydesdorff2011}).

\rui{} sits in the textual, citation-free family but differs from these measures
on two axes. First, unit: novelty is scored per paper and must be aggregated to
describe an institution, a step in which a few extreme papers dominate; \rui{}
instead compares one population of researchers against another, since a
university is a set of researchers, not of papers, and it is researchers who
produce papers. Second, reference: diversity is computed within a unit, whereas
\rui{} measures displacement relative to an external same-field standard
population. \emph{Uniqueness}, as we operationalise it, is thus a
reference-relative, population-level, citation-free property, distinct from
paper-level novelty and within-unit diversity. The constructs can move
independently: a unit may be concentrated yet typical, or diverse yet displaced.
We return to their empirical relations in Section~\ref{sec:diversity}.

\subsection{Researcher representation and optimal transport}
Two methods underpin \rui{}. For \emph{researcher representation} we build on
Researcher2Vec \citep{mochihashi2023}, which represents a researcher as a vector
derived from the text of their papers (originally for funding-agency
recommendation); we adopt it and give the construction in
Section~\ref{sec:methods}. To \emph{compare populations} we use Optimal Transport
\citep{villani2009,peyre2019} and its unbalanced variant \citep{chizat2018}, a geometry-aware way to compare
distributions of points that allows mass to be created or destroyed, with
entropic regularisation for tractability
\citep{cuturi2013,flamary2021}; our application to two embedding populations
follows \citet{kishino2025}, who track lexical semantic shift with unbalanced OT,
adapted here from words to researchers. Finally, the micro-level output of
\rui{} is a research-niche map (Section~\ref{sec:terms}), which connects it to
science mapping from citation links \citep{boyack2010,vaneck2010} or, increasingly,
document-level neural embeddings \citep{cohan2020}; \rui{} differs in that it
first quantifies displacement from an explicit standard population and returns a
comparable scalar, then maps.

\section{Data}
\label{sec:data}
We use bibliographic metadata and abstracts from Scopus (Elsevier), retrieved
through the Scopus APIs \citep{baas2020} between June and July 2026, restricted to
the neuroscience subject area (\texttt{SUBJAREA(NEUR)}) and to publication years
2021--2025. Scopus subject areas are its broad disciplinary categories; fixing
one is necessary because uniqueness is defined only against a same-field norm. We
use neuroscience as the demonstration field for our team's domain expertise---the
method itself is field-agnostic, and Section~\ref{sec:cs} replicates it in
computer science and economics---and the 2021--2025 window captures each university's current
research. A \emph{group} is a set of researchers we compare: the standard
population, each of the ten universities, and (for a construct-validity test)
five specialised institutes. Every researcher is represented only by their
2021--2025 neuroscience abstracts.

\paragraph{The standard population.}
We scanned \texttt{SUBJAREA(NEUR) AND PUBYEAR 2021--2025} for neuroscience author
identifiers and drew a random sample of 10{,}000; 9{,}861 had at least one
retrievable abstract and form the reference population. This is past the point of
convergence and robust to reference composition (Supplementary Section~S3).

\paragraph{Universities.}
For each university we retrieved its authors (same field and window) using the
main institutional Affiliation ID together with the affiliated medical school /
hospital ID, to avoid under-counting clinical neuroscience; Affiliation IDs and
per-university author counts are in Supplementary Table~S1. A researcher is
included on the basis of a single retrievable abstract, a deliberately low bar
that treats early-career and senior researchers alike (each contributes one unit,
regardless of output). The ten universities span North America (Harvard,
Stanford, MIT), the United Kingdom (Oxford, Cambridge), East Asia (Tsinghua,
Peking, the University of Tokyo, Kyoto), and Singapore (NUS).

\paragraph{Specialised institutes (construct-validity test).}
We additionally retrieved five dedicated, highly specialised neuroscience
institutes by their own Affiliation IDs (the Allen Institute for Brain Science,
HHMI Janelia, the Champalimaud Centre, Cold Spring Harbor Laboratory, and RIKEN
CBS; 190--390 researchers each), to test whether \rui{} ranks acknowledged
specialists above broad universities. They are not part of the ten-university
comparison.

\section{Methods}
\label{sec:methods}

\subsection{Researcher representation (Researcher2Vec)}
\label{sec:r2v}
We follow the Researcher2Vec framework \citep{mochihashi2023}, in which a
researcher is represented by a single vector aggregated from the text of their
papers. Let $\mathcal{P}$ be the set of all papers and $\mathcal{V}$ the whole
vocabulary.

\paragraph{Paper embeddings.}
From the paper--term count matrix
$X\in\mathbb{R}^{|\mathcal{P}|\times|\mathcal{V}|}$ (tokens of $\ge 4$ letters,
English stop words removed, document frequency in $[10,\,0.7\,|\mathcal{P}|]$)
we form the positive pointwise mutual information (PPMI) matrix
\citep{levy2014}:
\begin{equation}
M_{iv}=\max\!\Big(\log\frac{p(i,v)}{p(i)p(v)},\,0\Big),
\ \text{where}\ \ p(i,v)=\frac{X_{iv}}{\sum_{i,v}X_{iv}},\;\;
p(i)=\sum_{v}p(i,v),\;\; p(v)=\sum_{i}p(i,v)
\end{equation}
and reduce it with truncated SVD, $M\approx U\Sigma W^{\!\top}$, to yield a
$d$-dimensional embedding of the $i$th paper $\mathbf{x}_i=(U\Sigma)_i$.

\paragraph{Researcher vectors.}
Each researcher $r$ with a set of papers $\mathcal{P}_r$ is represented by the
mean of their paper embeddings,
\begin{equation}
\mathbf{v}(r)=\frac{1}{|\mathcal{P}_r|}\sum_{i\in\mathcal{P}_r}\mathbf{x}_i
\ \in\ \mathbb{R}^{d},
\end{equation}
the linear (averaging) instantiation of the Researcher2Vec representation. A
university $u$ and the standard then become corresponding point sets
$\mathcal{U}=\{\mathbf{v}(r):r\in u\}$ and
$\mathcal{S}=\{\mathbf{v}(r'):r'\in\text{standard}\}$, and distances are measured
by cosine.

Two points about scale follow from this construction. The researcher vectors are
not normalised, and need not be: the transport cost is a cosine distance, which
is invariant to the length of each vector, so normalising $\mathbf{v}(r)$ would
leave every score unchanged. The paper embeddings inside the average are not
normalised either, and that is a substantive choice: papers enter the mean in
proportion to their norm, so a lexically rich abstract pulls a researcher's
representation somewhat harder than a sparse one. Averaging unit-length paper
vectors instead---one vote per paper---leaves the institutional ranking
essentially unchanged (Supplementary Section~S4), so we report the simpler
unnormalised mean.

\paragraph{Choice of embedding.}
We deliberately use PPMI+SVD rather than a large contextual encoder. It is
transparent, deterministic, and cheap to reproduce, with no GPU or pretrained
weights, which matters for an evaluation indicator that others must be able to
recompute and audit \citep{hicks2015}; the representation follows Researcher2Vec
\citep{mochihashi2023}. Because \rui{} depends on the population-level geometry
of these vectors rather than on any single representation, its conclusions are
stable to the embedding: an independent TF-IDF embedding reproduces a similar
ranking (Section~\ref{sec:convergent}), and the mean aggregation is likewise not
load-bearing, the robust geometric median giving essentially the same
institutional picture (Supplementary Section~S4). We avoid citation-trained
encoders such as SPECTER \citep{cohan2020}, which would reintroduce the citation
signal \rui{} is designed to exclude.

\subsection{Uniqueness via unbalanced optimal transport}
\begin{figure}[t]
\centering
\includegraphics[width=\linewidth]{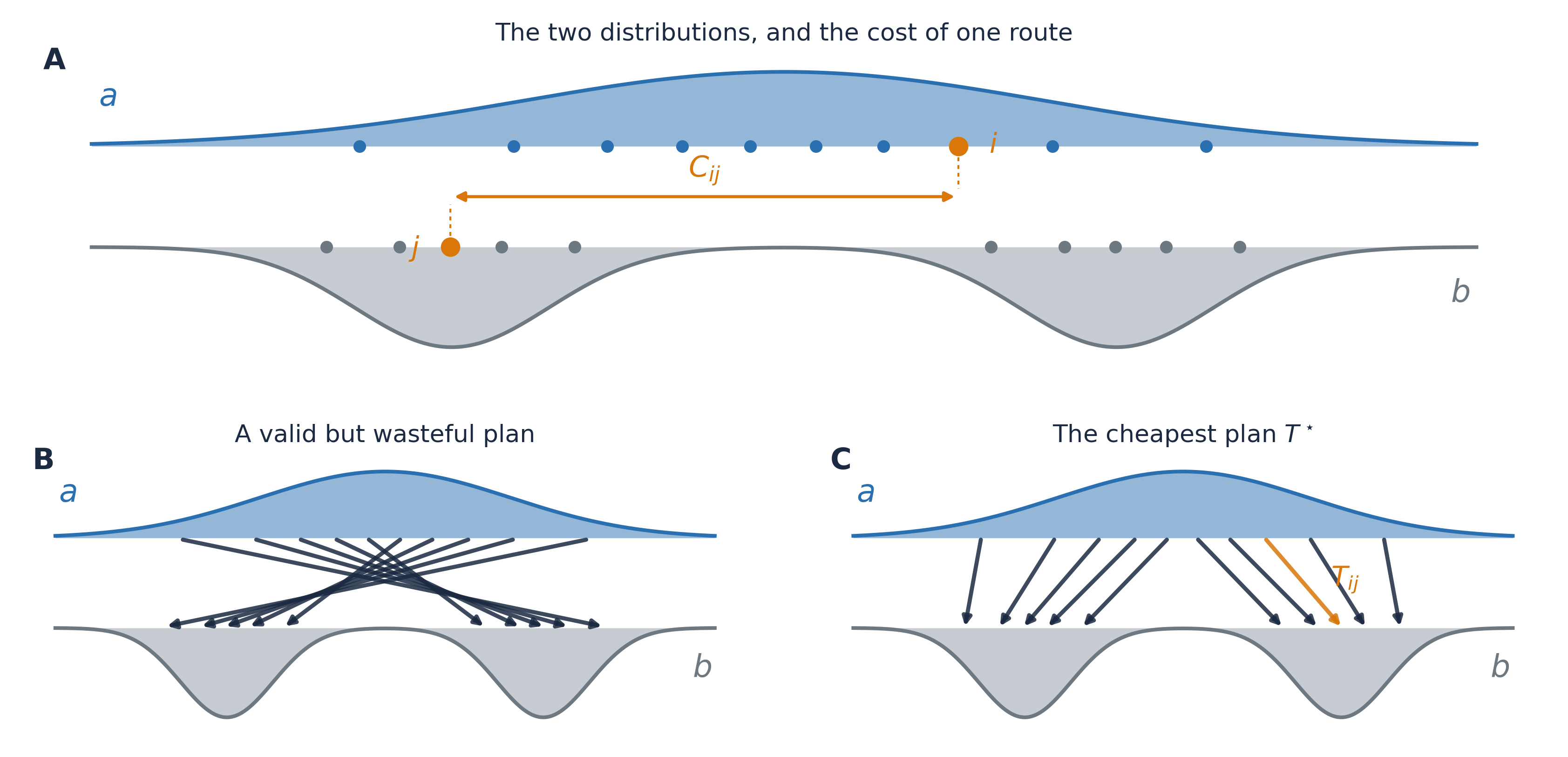}
\caption{What optimal transport does. Both $a$ and $b$ are probability
distributions over the same space, so each carries a total mass of one. That
space is drawn here as a single horizontal axis, because the space used in this
paper has far more dimensions than can be put on a page. \textbf{(A)}~Each
distribution is represented by ten points of equal mass, a number chosen only
to keep the drawing readable, and $b$ is mirrored below the axis. The cost
$C_{ij}$ of a route is how far the mass has to travel, drawn in orange between
the two marked points. \textbf{(B)}~Each arrow is one route of a plan $T$, and
this plan sends the mass across itself. \textbf{(C)}~The cheapest plan pairs
the two distributions in order, so that no two routes cross, and its total cost
$\langle T,C\rangle$ is about a quarter of that of the plan in B.}
\label{fig:otprimer}
\end{figure}

Optimal transport compares two probability distributions by asking what it
would cost to turn one into the other. Figure~\ref{fig:otprimer} sets out the ingredients on
a one-dimensional illustration. One distribution supplies mass and the other
demands it, a cost $C_{ij}$ says how expensive it is to move one unit from
point $i$ to point $j$, and a transport plan $T$ records how much mass travels
on each route, written $T_{ij}$. A plan is valid when its row sums return the
supply and its column sums return the demand, so that everything available is
sent and everything required is filled. Many plans meet those two conditions,
and optimal transport returns the one whose total cost
$\langle T,C\rangle=\sum_{ij}T_{ij}C_{ij}$ is smallest. What the method yields
is therefore not a single distance but an assignment, and it is the assignment
that carries the information we use.

In our setting the two distributions are populations of researchers in the same
embedding space. Let university $u$ have $n=|\mathcal{U}|$ researchers with
uniform weights $a=\tfrac{1}{n}\mathbf{1}$ and the standard population
$m=|\mathcal{S}|$ with $b=\tfrac{1}{m}\mathbf{1}$. Every researcher carries the
same mass whatever their output, so the index does not reward volume. The cost
$C_{ij}=1-\cos\!\big(\mathbf{v}_i,\mathbf{v}_j\big)$ is the cosine distance
between two researcher vectors, which compares the direction of a researcher's
work and sets aside the length of the vector, a quantity that reflects how many
abstracts were averaged into it rather than what they are about.

Classical optimal transport requires the row and column sums to match $a$ and
$b$ exactly, and that is the wrong requirement here. It forces every researcher
to be matched into the standard population however far the nearest counterpart
lies, so a researcher who belongs to no established community looks no
different from one at the centre of a crowd. The unbalanced formulation of
\citet{chizat2018} replaces the two equality constraints with penalties, and we
use a single penalty weight $\lambda$ on both sides,
\begin{equation}
T^\star=\arg\min_{T\ge 0}\ \langle T,C\rangle
-\varepsilon\,\mathrm{H}(T)
+\lambda\,\mathrm{KL}(T\mathbf{1}\,\|\,a)
+\lambda\,\mathrm{KL}(T^{\!\top}\mathbf{1}\,\|\,b).
\end{equation}
Here $T\mathbf{1}$ and $T^{\!\top}\mathbf{1}$ are the row and column sums of
the plan, that is, the mass actually sent and the mass actually received. The
Kullback--Leibler divergence $\mathrm{KL}$ measures how far those realised
totals fall short of the intended $a$ and $b$, and $\lambda$ is the price of
falling short. A small $\lambda$ makes giving up cheap and leaves most mass
unmatched, whereas a large one restores the balanced problem. The remaining
term uses the entropy $\mathrm{H}(T)=-\sum_{ij}T_{ij}(\log T_{ij}-1)$ and
follows \citet{peyre2019}. Subtracting a multiple of the entropy makes the
objective strictly convex and allows the solution to be found by the Sinkhorn
algorithm \citep{cuturi2013,chizat2018,flamary2021}, at the cost of spreading
each researcher's mass over several nearby counterparts rather than
concentrating it on one. Our formulation adapts the unbalanced-OT comparison of
\citet{kishino2025} from lexical to researcher embeddings.

\begin{figure}[t]
\centering
\includegraphics[width=\linewidth]{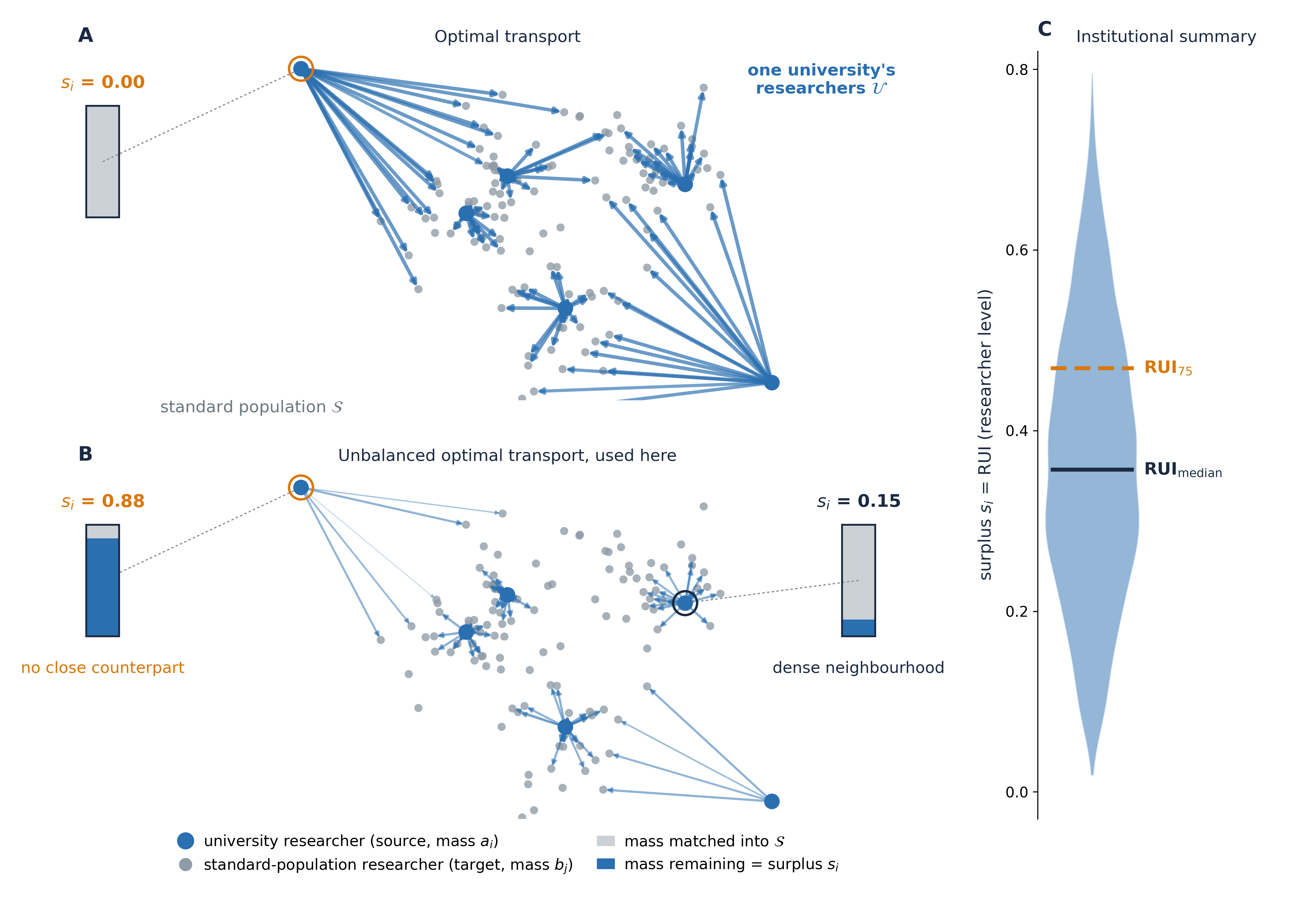}
\caption{How the unbalanced problem makes uniqueness visible. Every researcher
is one point in the $d$-dimensional embedding space, shown here as a
two-dimensional illustration. Blue points are one university's researchers
$\mathcal{U}$, which supply mass $a_i$, and grey points are the standard
population $\mathcal{S}$ from the same field, which demands mass $b_j$. Arrows point in the direction of transport and their width
is the amount delivered on that link. Each bar starts full and drains from the
top as mass is matched, so the blue level remaining is the surplus $s_i$. The
researcher circled in orange is the same one in both panels.
\textbf{(A)}~Under classical transport \citep{villani2009,peyre2019} every
researcher must send all of their mass into $\mathcal{S}$ however far it has to
travel, so an isolated researcher is indistinguishable from a mainstream one
and $s_i=0$ for all. \textbf{(B)}~The unbalanced problem used here
\citep{chizat2018} adds the penalty $\lambda$, which allows mass to be left
unmatched. The circled researcher has no counterpart nearby, giving up is
cheaper than transporting, and most of the mass stays. For a researcher inside
a dense part of $\mathcal{S}$ almost all of it finds close counterparts and
little remains. The entropic term spreads each plan over several neighbours
rather than concentrating it on one. \textbf{(C)}~Every researcher of a
university receives a score $s_i\in[0,1]$, and the institution is the
distribution of those scores, summarised by $\ruimed$ and $\rui_{75}$. Both
transport plans, and the example values of $s_i$, are actual solutions computed
on two-dimensional toy data with the paper's $\lambda=0.2$.}
\label{fig:uot}
\end{figure}

Because the penalty allows mass to be left behind, a researcher with no close
counterpart in $\mathcal{S}$ retains unmatched mass. For researcher $i$ the
\emph{surplus mass} is
\begin{equation}
s_i=\max\!\left(\dfrac{a_i-\sum_j T^\star_{ij}}{a_i},\,0\right)\in[0,1],
\end{equation}
the fraction of the researcher's mass that cannot be matched into the standard
population. Figure~\ref{fig:uot} shows the mechanism on a two-dimensional
illustration. The researcher who is forced into a distant match under balanced
transport keeps most of their mass once the constraint becomes a penalty, while
a researcher inside a dense neighbourhood keeps almost none. One consequence of
the entropic term is that two identical populations do not reach a surplus of
zero, so we read $\ruimed$ against the value the standard population obtains
against itself rather than against zero.

 Aggregated over a unit, the surplus behaves like a
regularised excess-mass divergence $\int\max(q-p,0)$ between the unit's
empirical density $q$ and the standard's $p$; because $p$ is spread across the
whole field, this divergence is larger for units displaced into sparsely
populated regions of the space, the geometric basis of the construct we probe
in Section~\ref{sec:diversity}. We call $s_i$ researcher $i$'s \rui{}: it is the
index at its native, per-researcher level, and the two institutional figures
below are summaries of its distribution over a unit. The headline institutional
value is the median,
\begin{equation}
\ruimed(u)=\operatorname*{median}_{i\in u}\, s_i .
\end{equation}
We also report its upper quartile,
\begin{equation}
\rui_{75}(u)=Q_{0.75}\{\,s_i : i\in u\,\},
\end{equation}
the 75th percentile of the per-researcher \rui{}. In one line: $\ruimed$ is the
uniqueness of the \emph{typical} researcher, and $\rui_{75}$ is the uniqueness of
a university's \emph{most unique} researchers: the level reached by its top
quarter (analogous to the substantiality construct of \citealp{shirabe2021}). We
read $\ruimed$ to compare a unit's overall level and $\rui_{75}$ for its most
unique researchers (Section~\ref{sec:subst}). Throughout, an unsubscripted
\rui{} denotes either the index in general or a researcher's own score; a
subscript is used whenever an institutional value is meant. For interpretation,
$\ruimed(u)$ can also be read as a percentile within the standard population's
own surplus distribution (standard-vs-standard).

\subsection{Confirmed configuration}
All analyses use \texttt{dim}$=40$, $\lambda=0.2$, $\varepsilon=0.005$, and
\texttt{min\_papers}$=1$ (all researchers embedded). These settings maximise the
entropy (spread) of the per-researcher score distribution, so that the
configuration separates researchers as sharply as possible rather than bunching
them at similar scores; the entropy criterion, the grid, and its corroboration
(agreement with the independent-embedding ranking and negligible resampling
dispersion) are given in Supplementary Sections~S5--S7. Two displays in
Section~\ref{sec:terms} additionally restrict to researchers with $\ge 3$
neuroscience papers, to avoid the degenerate ties produced by single-paper
authors: the characteristic-term table, and the niche map, which further selects
its top researchers by an output-adjusted score (Section~\ref{sec:terms}). These
are presentation choices only and do not affect the ranking.

\section{Results}
\label{sec:results}

\subsection{Uniqueness ranking}
Our results should be read at three levels, which the evidence supports to
different degrees. \emph{Robust:} every university sits well above the standard,
a gap that survives every check we ran. \emph{Interpretable:} the per-researcher
scores identify who is unique and in what, which is where the index is most
informative. \emph{Qualified:} the order among adjacent universities,
which is sensitive to resampling, productivity and institutional boundaries. We
present the ranking because it is the natural summary of the index, but we
interpret only the first two levels.

\begin{figure}[t]
\centering
\includegraphics[width=\linewidth]{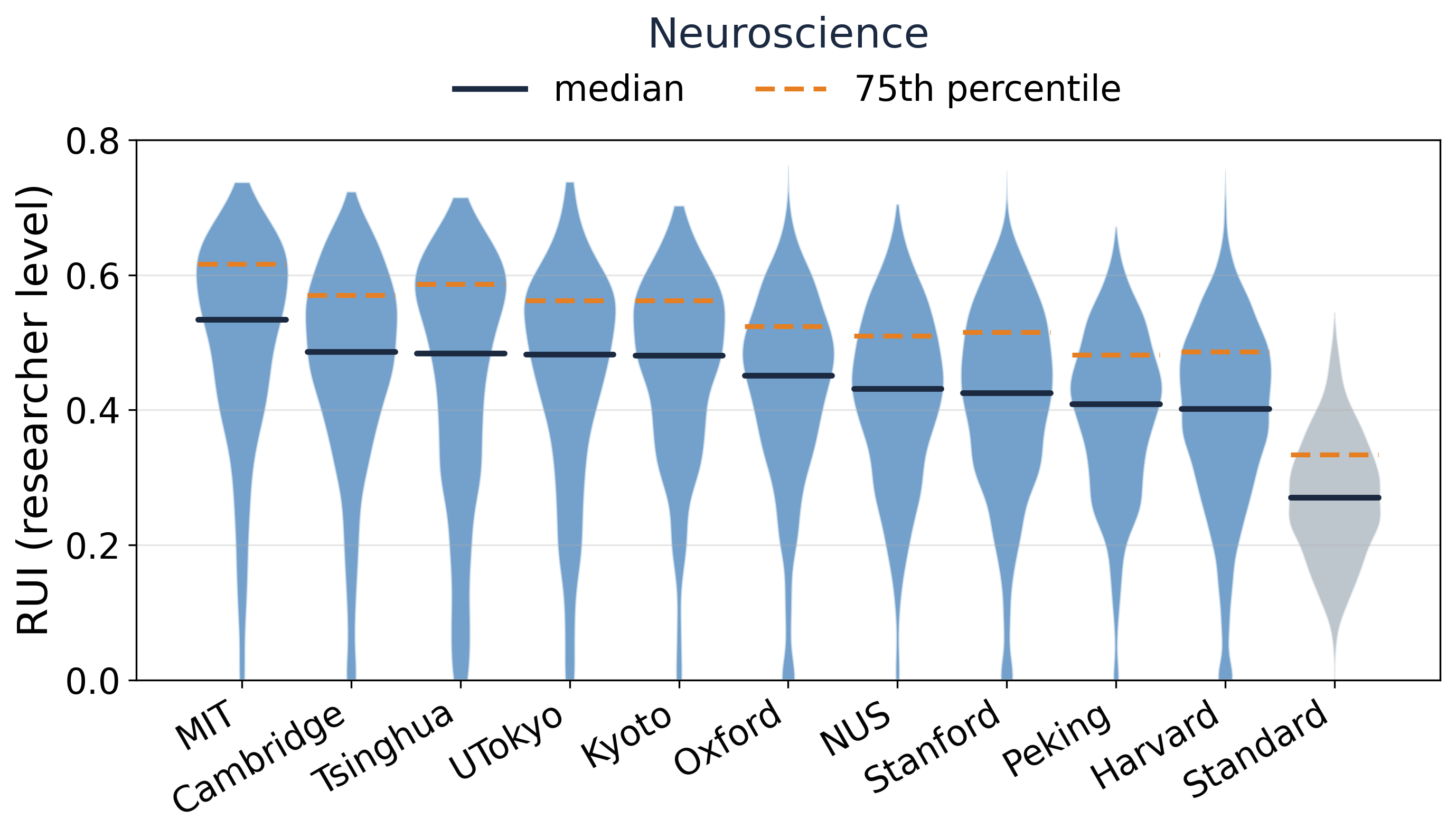}
\caption{Distribution of per-researcher uniqueness (surplus mass) by
university versus the standard, ordered by median. The width of each violin at a
given height is proportional to the density of researchers at that level of
uniqueness (wider means more researchers there); each violin is a kernel density
estimate over that university's researchers. The solid navy line marks the
median (the university's $\ruimed$) and the orange dashed line the 75th percentile
($\rui_{75}$, Section~\ref{sec:subst}).}
\label{fig:violin}
\end{figure}

Table~\ref{tab:ranking} and Figure~\ref{fig:violin} show that all ten
universities are clearly more unique than the standard. We rank universities by
their $\ruimed$, the median per-researcher \rui{}: it runs 0.40--0.53 across the
ten, against 0.27 for the standard, so every university's typical researcher is
more unique than $\sim$92--99\% of standard-population researchers. This gap,
every university well above the standard, is the robust, design-insensitive
result, and it is what we interpret. The
order among the universities, by contrast, should be read with caution:
adjacent universities differ by less than the resampling variability, the
per-researcher score depends on productivity (Section~\ref{sec:prod}), the
institutional scalar is largely reducible to topic concentration
(Section~\ref{sec:diversity}), and the ranking is sensitive to how each
institution's boundary is drawn, which for a multi-institution field like
neuroscience is hard to avoid (Supplementary Section~S2). We
therefore read the ranking, under the main-plus-medical affiliation definition,
with care: its extremes and broad clusters are stable, whereas the precise
order among adjacent universities is not. The index is not intended as a league
table.

\begin{table}[t]
\centering
\caption{University research uniqueness (\rui{}) in neuroscience,
2021--2025. \texttt{dim}$=40$, $\lambda=0.2$, $\varepsilon=0.005$,
\texttt{min\_papers}$=1$. Values are the median and 75th percentile of the
per-researcher surplus mass ($\ruimed$ and $\rui_{75}$); the 75th percentile is the
level exceeded by a university's most unique quarter of researchers
(Section~\ref{sec:subst}). Universities are ordered by $\ruimed$ for
presentation; the index is not intended as a league table.}
\label{tab:ranking}
\begin{tabular}{lcccc}
\toprule
University & Country & $n$ & $\ruimed$ & $\rui_{75}$ \\
\midrule
MIT & USA & 1{,}174 & 0.534 & 0.616 \\
Cambridge & UK & 2{,}059 & 0.486 & 0.570 \\
Tsinghua & China & 1{,}935 & 0.484 & 0.587 \\
Univ.\ Tokyo & Japan & 1{,}596 & 0.483 & 0.563 \\
Kyoto & Japan & 1{,}095 & 0.481 & 0.562 \\
Oxford & UK & 3{,}564 & 0.451 & 0.524 \\
NUS & Singapore & 1{,}509 & 0.432 & 0.510 \\
Stanford & USA & 4{,}982 & 0.425 & 0.515 \\
Peking & China & 3{,}295 & 0.409 & 0.482 \\
Harvard & USA & 9{,}306 & 0.402 & 0.487 \\
\midrule
Standard & -- & 9{,}861 & 0.271 & 0.334 \\
\bottomrule
\end{tabular}
\end{table}

\subsection{Two readings of the distribution: the typical researcher and the
most unique quarter}
\label{sec:subst}
Because \rui{} is a whole distribution over researchers (Figure~\ref{fig:violin}),
not a single number, it supports two complementary summaries. The median
($\ruimed$) is the uniqueness of the \emph{typical} researcher; the upper quartile
($\rui_{75}$) is the level reached by the \emph{most unique quarter}, that is,
how far uniqueness extends among a university's most unique researchers. The choice
of the 75th percentile has direct precedent in research evaluation: THE's World
University Rankings define ``Research Strength'' as the 75th-percentile
field-weighted citation impact \citep{the2024wur}, chosen because it resists a
few extreme papers. $\rui_{75}$ is a citation-free, per-researcher counterpart,
and a distributional reading of the substantiality construct of
\citet{shirabe2021} (the accumulation of research excellence), though as a
size-independent level rather than an accumulation count. Like the median, it
uses no citations.

The two readings need not agree, and where they diverge is itself informative. In
neuroscience they almost coincide (rank correlation $\rho=0.96$): a university
whose typical researcher is unique also has a strongly unique top quarter. In other
fields they separate---in computer science and economics the two orderings even
invert at the top---so $\rui_{75}$ is a second, distribution-aware view of the same
index rather than a rival league table, distinguishing universities that are
uniformly unique from those whose uniqueness is concentrated in a small group of
standout researchers. The cross-field comparison, and a check that the inversions
are not a productivity artefact, are in Supplementary Section~S9.

The ranking is stable: over 20 random halves of the standard the medians vary by
$\le 0.003$ and the rankings agree with Kendall's $W=0.96$, and an independent,
disjoint re-draw of the standard reproduces every university's $\ruimed$ to within $0.005$
(Supplementary Section~S10).
\label{sec:stability}
Beyond how much each university departs from the standard, the same per-researcher
scores reveal \emph{what} it departs in---which researchers are most unique, and
on which topics---to which we turn next.

\subsection{Research fingerprints: characteristic terms and niches}
\label{sec:terms}
This is where the micro view pays off: below the institutional score, each
university leaves a recognisable fingerprint. For the most unique
researchers of each university (selected, as everywhere in this section, by the
output-adjusted residual) we extract characteristic terms by normalised
pointwise mutual information \citep{bouma2009} between researchers (treated as
documents) and terms, over the same topic-specific vocabulary that labels the
fingerprints below. The recovered niches are vivid, specific, and rarely
overlap: language and moral cognition (MIT); diffusion-MRI fibre dispersion and
visual perception (Oxford); care-home SARS-CoV-2 transmission, dominated by a
single laboratory (Cambridge); brain--computer interfaces (Tsinghua);
\emph{C.\ elegans} salt chemotaxis and dendritic-spine imaging (Tokyo);
blindsight and mesial-temporal epilepsy (Kyoto); cortical parcellation and
cross-site harmonisation (NUS); retinal prostheses and immuno-proteomics
(Stanford); calcium-imaging optics and neuropeptide sensors (Peking); fetal
white-matter tractography and narcolepsy (Harvard). Full lists are in
Supplementary Table~S22.

\begin{figure}[t]
\centering
\includegraphics[width=0.8\textwidth]{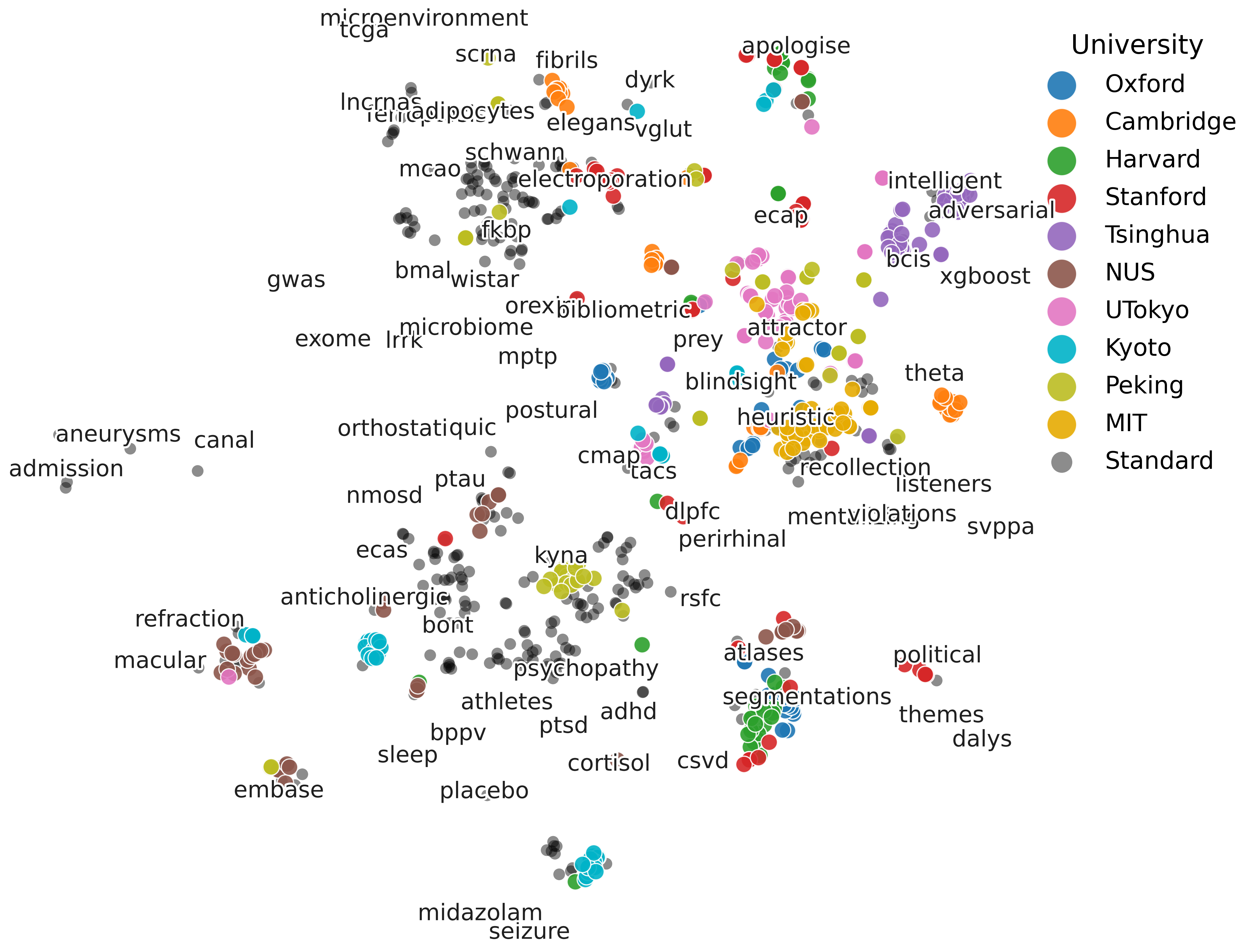}
\caption{Research-niche map. The 50 most unique researchers of each university
(by output-adjusted uniqueness; see text) and the most topic-specific field
terms are embedded together in two dimensions with t-SNE; points are coloured by
university and grey words label the regions. Terms are selected by their
concentration across latent (LDA) topics, which removes field-generic vocabulary
and keeps only topic-characteristic words.}
\label{fig:niche}
\end{figure}

\label{sec:niche}
We can also read this as a map. We embed the most unique researchers (the top
50 per university, ranked by uniqueness adjusted for output; see below) together
with the terms that most sharply
characterise the field's topics, in a single two-dimensional layout, so that the
meaning of each region can be read directly from the words that fall in it
(Figure~\ref{fig:niche}). Each term is selected by how concentrated it is across
latent topics, so field-generic vocabulary drops out and only topic-specific
words remain. Universities occupy recognisable, field-coherent regions:
brain segmentation and small-vessel imaging (Oxford and Harvard, among terms such
as segmentations, atlases and csvd); machine-learning and brain-computer
interfaces (Tsinghua, among python, adversarial, xgboost and bcis); semantic
memory and affect (MIT, among heuristic, recollection, disgust and listeners);
attractor dynamics and olfactory behaviour (the University of Tokyo, among
attractor and prey); ophthalmology (NUS, among myopia and macular); and epilepsy
together with blindsight (Kyoto, among seizure, perampanel and ccep). Read together with the characteristic terms above, the map shows
that each university's signature is less a uniform distinctiveness than the
imprint of a few unusually distinctive groups.

\paragraph{How the map is built.}
We fit a twenty-topic latent Dirichlet allocation---the same topic count used
for the university--topic fingerprints below---and score each vocabulary term by
how concentrated it is across topics, $\log T - H(p(t\mid w))$, where
$p(t\mid w)$ is the term's posterior distribution over the $T$ topics. Terms that
spread evenly across topics (field-generic words) score near zero and are
dropped, whereas topic-specific terms score highest. The ranking is automatic,
but it is applied on top of a stop-word list that extends the standard English
list with generic scientific and editorial vocabulary; the same list is used for
the characteristic terms and the fingerprints, so all three displays of this
section share one vocabulary rule. The $1{,}000$ top-scoring terms are
represented by their factors in the same PPMI+SVD space as the papers and are
embedded, in a single t-SNE, together with the top-50 most unique researchers of
each university, so that researchers and the terms that describe them fall near
one another. To keep the map readable, overlapping labels are thinned before
drawing, so $75$ of those $1{,}000$ terms are actually printed; the omitted terms
are those whose position collides with an already-drawn label. The 50 researchers are selected not by raw
surplus but by an output-adjusted score: on the standard population we fit a
simple regression of surplus on the log number of papers, $g(n)=a+b\log n$ (with
$b<0$, since surplus declines as output grows), and rank each researcher by the
residual $s_i - g(n_i)$, that is, how much more unique they are than a standard
researcher with the same output. This keeps the map from being dominated by
low-output researchers, whose mean vector is mechanically off-centre; the niche
regions are the same as under raw selection, but the selected researchers are
more representative.

\begin{figure}[t]
\centering
\includegraphics[width=\linewidth]{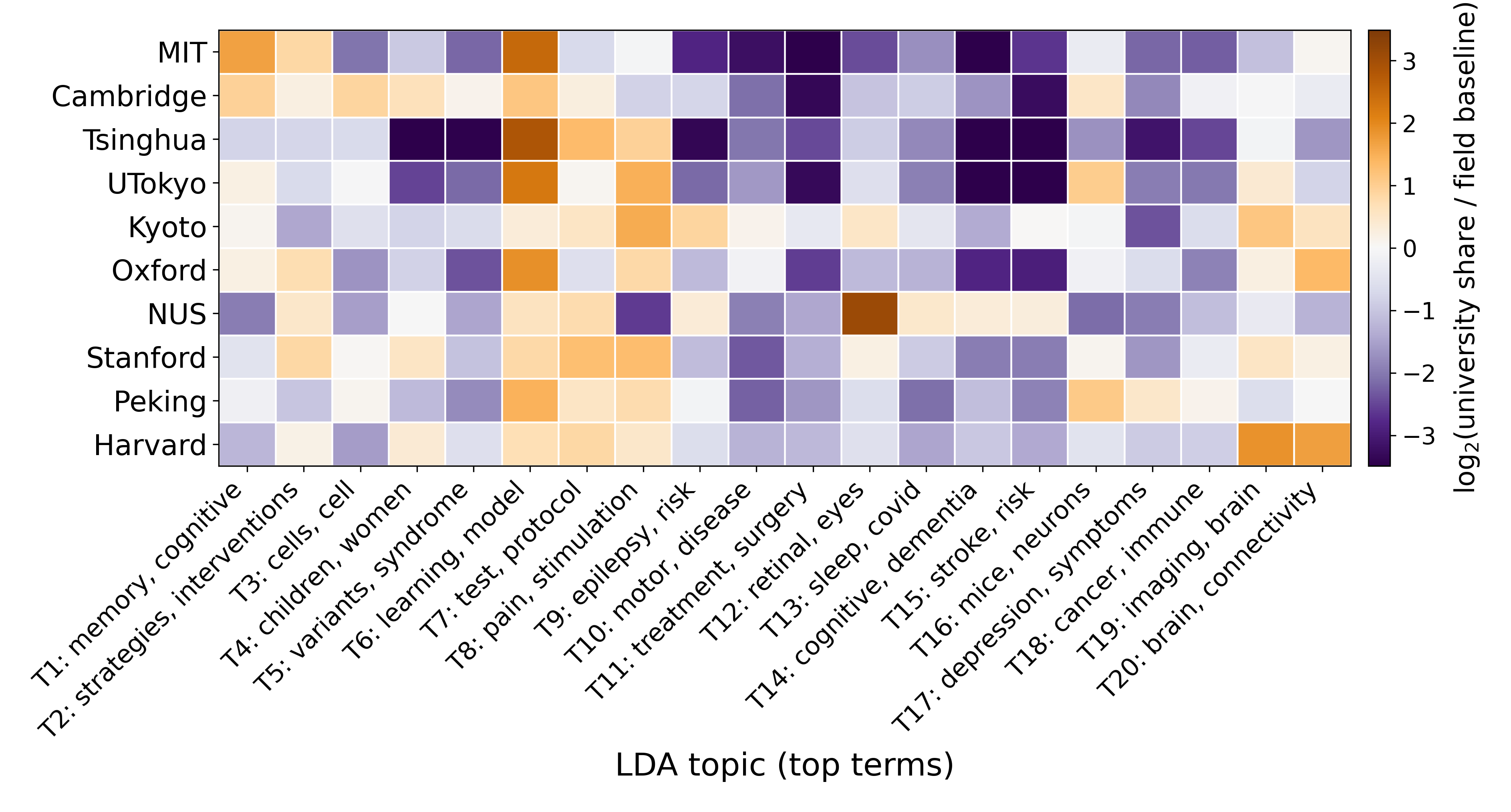}
\caption{University--topic fingerprints. Each row is a university and each column
a latent (LDA) topic, labelled by its top terms. Colour is the $\log_2$ ratio of
the topic's share among the university's 200 most unique researchers
(output-adjusted) to its share in the standard population: orange = over-,
purple = under-represented relative to the field. Columns (topics) are ordered by
the university that most over-represents each, and rows follow the $\ruimed$ ranking.}
\label{fig:unitopic}
\end{figure}

\paragraph{University--topic fingerprints.}
The niche map places individual researchers; a complementary view aggregates
them into a compact institution-level summary. We fit a twenty-topic latent
Dirichlet allocation to the standard population's abstracts, which gives a stable
set of field topics (cellular and synaptic physiology, epilepsy, stroke, dementia,
functional connectivity, structural imaging, computational models, and so on;
generic and editorial vocabulary is removed with an extended stop-word list). For
each university we pool the papers of its 200 most unique researchers (the same
output-adjusted selection, broadened from 50 to stabilise the topic estimates)
and average their topic distributions, then compare this profile with the field
baseline. Figure~\ref{fig:unitopic} plots the log-ratio: orange marks topics that
a university's most unique researchers over-represent relative to the field,
purple those they under-represent. Two patterns stand out. First, a shared
signature: at almost every leading university the most unique researchers lean
towards computational and machine-learning work relative to the clinical-heavy
field, most sharply at Tsinghua, MIT and Tokyo. Second, clear institution-specific
niches on top of it---ophthalmology at NUS (by far its strongest
over-representation), structural and functional neuroimaging at Harvard,
neurostimulation and imaging at Kyoto, and cellular/molecular neuroscience at
Tokyo and Peking. These give a coarser, institution-level companion to the
per-researcher niche map.

\subsection{Construct validity}
\label{sec:diversity}
Is \rui{} merely a restatement of an existing indicator? Across the ten
universities it is not significantly correlated with Rao--Stirling diversity
(Spearman $\rho=-0.47$, n.s.) or with the naive distance from each university to
the standard, to its centroid ($\rho=0.47$) or to the ten nearest standard
researchers ($\rho=-0.10$; both n.s.), so it is neither a conventional
diversity index nor a simple distance. It is, however, strongly related to
topic concentration: institutions whose researchers spread evenly across
sub-fields (high Shannon entropy of their topic distribution) have low $\ruimed$,
and concentrated institutions have high $\ruimed$ ($\rho=-0.90$, $p<0.001$;
$\rho=-0.92$ after partialling out institution size, which is itself only
weakly related to entropy). This is expected rather than incidental: the
surplus mass behaves like a regularised excess-mass divergence
$\int\max(q-p,0)$ between a university's density $q$ and the standard's density
$p$, and because the standard is spread across the whole field, that divergence
is mechanically larger for institutions concentrated in distinctive niches.

Crucially, this reducibility is confined to the institutional scalar.
At the researcher level, topic membership (twenty $k$-means sub-fields) explains
only about a quarter of the variance in per-researcher surplus ($R^2=0.25$);
roughly three quarters is within-topic, so two researchers in the same sub-field
can receive very different scores. \rui{} thus adds little beyond niche
concentration when collapsed to one number per institution, but is not
reducible to a topic distribution where it is actually used, at the level of
individual researchers and the niche map (Figure~\ref{fig:niche}). The
institution-level correlations rest on $n=10$ and are read as indicative.

\paragraph{Displacement, not concentration.}
A controlled simulation separates the two: units held at near-zero topic entropy
but placed farther from the field's centre receive sharply higher \rui{}
($\rho=0.89$ with displacement), whereas a dispersed, high-entropy unit scores low
(Supplementary Section~S14, with figure). At the researcher level, adding the
distance to the standard centroid to a topic-only model raises $R^2$ only from
$0.25$ to $0.28$: most of the per-researcher signal is local transport geometry
(the unmatched-mass structure that distinguishes unbalanced OT from diversity and
naive distance), reproducible under an independent embedding
(Section~\ref{sec:convergent}). We therefore read \rui{} as a reference-relative
displacement of a unit's knowledge profile---related to, but not reducible to,
within-unit concentration or diversity.

\begin{figure}[t]
\centering
\includegraphics[width=\linewidth]{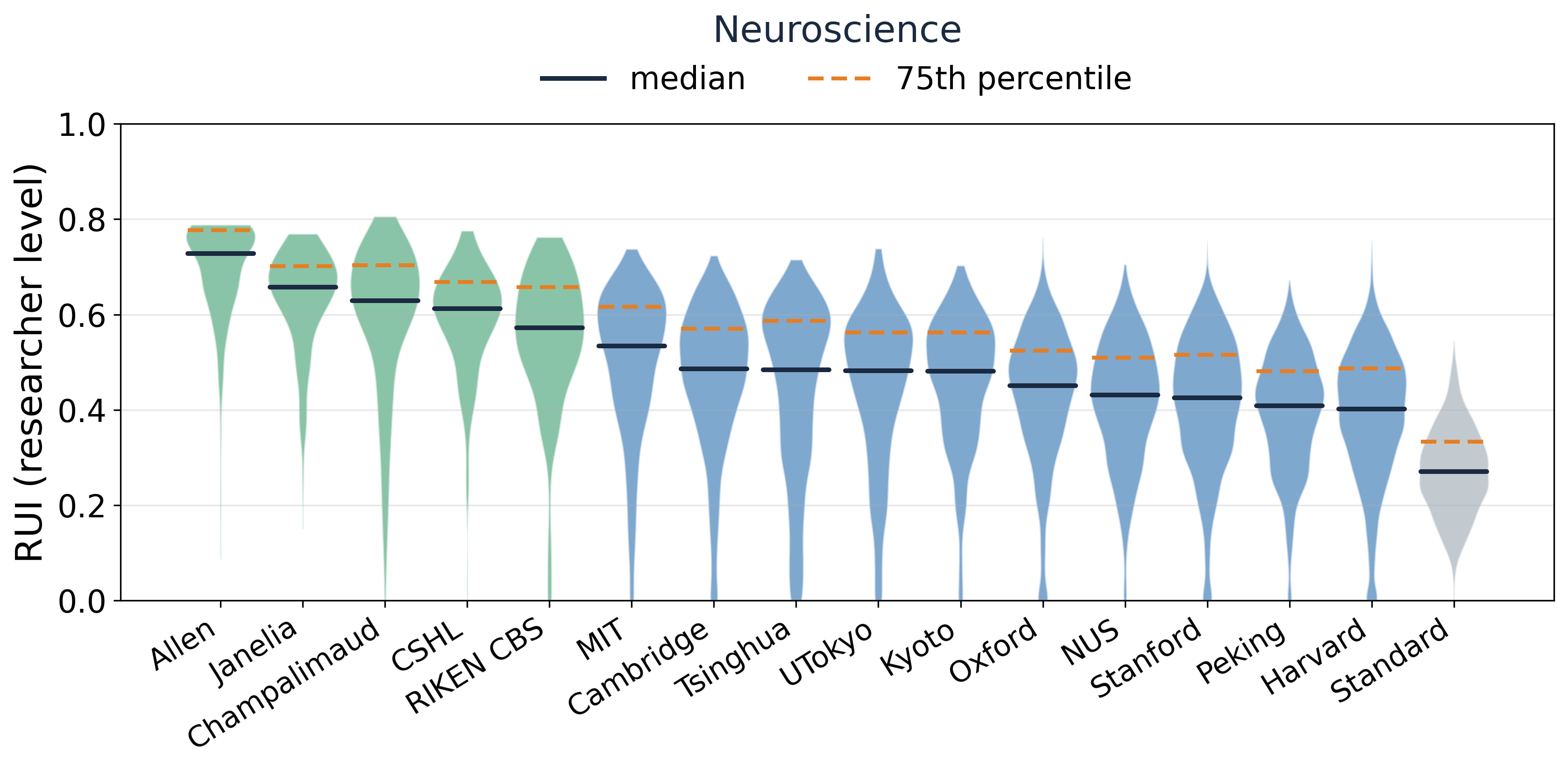}
\caption{Per-researcher uniqueness for the five specialised institutes (green),
the ten universities (blue), and the standard population (grey), ordered by
median within each group. The navy line marks the median and the orange dashed
line the 75th percentile. Every institute lies above every university.}
\label{fig:instviolin}
\end{figure}

\paragraph{Known-groups validation: specialised institutes rank above universities.}
\label{sec:institutes}
The checks above establish what \rui{} is not. A known-groups test asks the
positive question: does the index place institutions that are independently
recognised as distinctive above broad, general universities? We applied
the index, unchanged, to five independent and highly specialised neuroscience
institutes retrieved by their own Scopus affiliation IDs over the same field and
window: the Allen Institute for Brain Science, HHMI Janelia, the Champalimaud
Centre, Cold Spring Harbor Laboratory, and RIKEN CBS (190--390 researchers
each). All five rank above every university ($\ruimed$ $0.57$--$0.73$ versus
$0.40$--$0.53$ for the universities; Figure~\ref{fig:instviolin} and
Table~\ref{tab:institutes}).

This ordering is robust. Subsampling every unit to a common $n=190$ leaves the
institutes above all universities (institute minimum $0.58$ versus university
maximum $0.55$), even though the smaller samples raise universities'
surplus; bootstrap $95\%$ confidence intervals for the institutes do not overlap
those of any university (e.g.\ RIKEN CBS $[0.55,0.59]$ versus MIT, the most
unique university, $[0.52,0.54]$); and an independent TF-IDF embedding preserves
the ordering (rank correlation $0.99$; Supplementary Section~S15). The institutes are smaller and
topically concentrated, and they share collaborators with universities, which
would pull them towards the university mean, yet the gap persists. Institutions
selected in advance, on grounds entirely external to the index, land where a
valid uniqueness measure should place them. This is the paper's positive
criterion evidence, and it holds at the institutional level, where our other
external check does not.

\begin{table}[t]
\centering
\caption{Construct validity: five specialised neuroscience institutes, ranked by
$\ruimed$. All five exceed every university (cf.\ Table~\ref{tab:ranking};
the most unique university, MIT, is at $0.534$).}
\label{tab:institutes}
\begin{tabular}{llcc}
\toprule
Institute & Country & $n$ & $\ruimed$ \\
\midrule
Allen Institute for Brain Science & USA & 347 & 0.728 \\
HHMI Janelia & USA & 252 & 0.657 \\
Champalimaud Centre & Portugal & 194 & 0.629 \\
Cold Spring Harbor Laboratory & USA & 240 & 0.612 \\
RIKEN CBS & Japan & 388 & 0.572 \\
\bottomrule
\end{tabular}
\end{table}

\subsection{Validity and robustness checks}
\label{sec:convergent}
We summarise the checks here; full details are in the Supplementary Material.
\emph{Convergent and size validity.} An independent TF-IDF embedding reproduces
the ranking (Spearman $\rho=0.98$; same top three; Harvard lowest; Supplementary
Table~S18), and subsampling every university to a common $n=1{,}000$ researchers
preserves it ($\rho=0.98$; Supplementary Table~S12), so \rui{} is not an artefact
of the particular embedding or of institution size.

\paragraph{Discriminant validity: not impact, not novelty, not simple distance.}
\label{sec:novelty}
Per-researcher uniqueness is essentially uncorrelated with citation impact
(Spearman $\rho=-0.02$, $n=30{,}515$; Supplementary Table~S19), so \rui{} is not a
proxy for impact. At the institutional level, \rui{} and $\rui_{75}$ are likewise
uncorrelated with citation-based excellence indicators, both the institutional
h5-index (the representative substantiality metric of \citet{shirabe2021};
Spearman $\rho=-0.35$ and $-0.28$) and the field top-$10\%$ paper share
($\rho=+0.19$ and $+0.15$; Supplementary Section~S17), so uniqueness is distinct
from the accumulation of citation excellence, not only from average impact. It is also distinct from citation-based
novelty: against the
atypical-combination novelty of \citet{uzzi2013}, university researchers are more
novel than the standard at the population level, yet the two do not co-order
across universities (Supplementary Section~S18). Finally, \rui{} is not a
relabelled embedding-distance novelty. Compared with a distance-based novelty
computed in the same embedding space (each researcher's mean distance to their
nearest field neighbours, in the spirit of \citealp{shibayama2021}), the two
rank researchers very differently, sharing only about $22\%$ of each university's
top-50 (researcher-level Spearman $\rho=0.35$; Supplementary Section~S19). A
simple distance rewards isolated outliers, whereas \rui{} rewards systematic
displacement from the standard's density, so the two pick out different people.

\paragraph{Productivity confound.}
\label{sec:prod}
Per-researcher surplus declines with output (Spearman $\rho=-0.23$,
$n=30{,}515$), a mechanical effect of averaging rather than lower intrinsic
uniqueness. Controlling for it, by recomputing medians within paper-count strata
and against a paper-count-matched reference, leaves the cluster structure and the
extremes intact (Harvard lowest; $\rho=0.88$ with the original ranking), while
some middle ranks are output-dependent (Tsinghua in particular; Supplementary
Section~S12). Larger institutions also tend to have lower $\ruimed$, an association
that strengthens once the small, specialised institutes are included
(Supplementary Section~S13). We therefore interpret clusters and robust extremes
rather than a precise ordering of institutions.

\subsection{Generalization to further fields}
\label{sec:cs}
To test whether these findings are specific to neuroscience, we repeated the
entire pipeline in two further fields, computer science and economics (Scopus
subject areas \texttt{COMP} and \texttt{ECON}, 2021--2025), each with a fresh
random standard population of $\sim$10{,}000 same-field researchers and the same
ten universities, retrieved by the same affiliation IDs. The core result
replicates in both. In computer science all ten universities are markedly more
unique than the standard ($\ruimed$ $0.41$--$0.48$ versus $0.27$; lowest
standard-population percentile $92\%$); in economics the same holds, and even
more strongly ($\ruimed$ $0.50$--$0.57$ versus $0.32$; lowest percentile
$98\%$; per-field distributions and full rankings in
Supplementary Section~S21). The robust claim, that a university's research is
measurably displaced from what is typical in the field, therefore holds in two
further, methodologically distinct fields.

The ordering among universities, by contrast, is field-specific, which
reinforces our central caution that the gap above the standard is a stable
property whereas the rank among institutions is not.

\section{Discussion}
This paper proposes the Research Uniqueness Index (\rui{}), a citation-free
measure of how far a university's researchers depart, in knowledge structure,
from a same-field standard population. Each researcher is represented by a
PPMI+SVD embedding of their abstracts, and an unbalanced formulation of optimal
transport returns the share of that mass which cannot be matched into the
standard; the distribution of these per-researcher scores is the indicator,
summarised where necessary by quantiles. Applied to ten leading universities
against $9{,}861$ same-field researchers in neuroscience, all ten sit well above
the standard ($\ruimed$ $0.40$--$0.53$ versus $0.27$), and the result replicates
in computer science and economics (Section~\ref{sec:cs}). Construct validity is
supported by a known-groups test in which five independently recognised
specialised institutes rank above every broad university
(Section~\ref{sec:institutes}); discriminant validity by the absence of
correlation with citation impact; and robustness by resampling, equal-$n$
subsampling and an independent embedding. Two threads are worth drawing out.

\paragraph{What does uniqueness mean, and is it good?}
\rui{} measures displacement from what is typical in the field, not quality.
Distance
alone could signal pioneering work or merely peripheral, idiosyncratic work,
and our discriminant check is deliberately blunt about this: uniqueness is
essentially uncorrelated with citation impact, so the index is not a relabelled
measure of success. What keeps it from being mere eccentricity is the external evidence summarised
above---a margin over broad universities that survives size-matching,
bootstrapping and a change of embedding---together with a second, weaker signal:
the universities that \rui{} places well above the standard also make more of
the atypical reference combinations that, in prior work, foreshadow long-run
impact \citep{uzzi2013}. We therefore read \rui{} as a descriptive indicator of
where a community sits relative to its field, a complement to output and impact
metrics rather than a competitor to them. Methodologically it is a
size-independent, intensive indicator, in the family of field-weighted citation
impact and its 75th-percentile ``research strength'' variant \citep{the2024wur},
rather than the size-dependent accumulation indicators (publication or top-cited
counts, institutional h-indices) that grow with size; unlike all of these, it
uses no citations. It therefore sits beside quantity, impact and substantiality
\citep{shirabe2021} as a fourth dimension, derived from research content rather
than from how that content was received. This places \rui{} alongside
responsible-metrics initiatives such as the Leiden Manifesto \citep{hicks2015},
which caution against evaluation by counts alone: producing research unlike
anyone else's is arguably a core purpose of a university, and \rui{} measures it
without recourse to citations.

\paragraph{Who might use it.}
Different users would ask the index different questions. For a university's
research administrators and institutional-research offices, it identifies where
the institution's work is genuinely off-centre and which groups carry that
distinctiveness---information that volume- and citation-based reporting does not
surface. For funders and evaluation bodies, it offers a field-controlled
comparison that does not wait for citations to accumulate, and so can describe a
unit's position in its field years before impact measures stabilise. For a
department or faculty head, the per-researcher scores make it possible to argue
for supporting unusual work with more than intuition; and for researchers
themselves, the score says where their own work sits relative to their field,
which publication and citation counts do not report. In every case the use is
descriptive---portfolio diagnosis rather than league-table competition. Because
the reference is defined per field and no citations enter, the same construction
transfers across fields without the field-specific normalisation that
citation-based indicators require, and the entropy-based selection of dimension
and transport hyperparameters keeps these uses reproducible and free of ad-hoc
tuning.

\section{Limitations and future work}
The validity evidence is summarised at the start of the Discussion; full details
are in Section~\ref{sec:convergent} and Section~\ref{sec:novelty}. One
qualification is worth adding: the second external check, against
reference-based atypical-combination novelty, holds at the population level but
not as an inter-university ordering. The following caveats remain.

\subsection{Limitations}
The institutional ranking depends on where a university's boundary is drawn.
For a field like neuroscience, whose work is spread across universities, medical
schools, and hospitals, some sensitivity here is unavoidable: including or
excluding an affiliated medical school can change the order (Supplementary
Section~S2). Harvard is the clearest case, least unique with its
medical school included and most unique without it, because universities differ
in how their clinical research is organised and indexed in Scopus. The order
among universities therefore calls for care: the stable comparison is between a
university and its field, not between universities whose boundaries are drawn on
incomparable terms. The aggregate is also largely reducible to topic
concentration (Section~\ref{sec:diversity}), and within a fixed affiliation
definition the per-researcher score is confounded by productivity (the order
survives stratification and a paper-count-matched reference, but middle ranks
remain productivity-dependent; Section~\ref{sec:prod}).

Several data- and presentation-level caveats also apply. Our measure of the
standard is an approximate random sample drawn from a capped scan of the field;
an independent, disjoint re-draw reproduces the gap above the standard almost
exactly (Section~\ref{sec:stability}), but a formal representativeness analysis
of the underlying field scan remains desirable. Researchers are identified by
Scopus Author IDs, whose disambiguation is imperfect; split or merged identities
add noise to individual vectors, although the population-level gap above the
standard, an aggregate over thousands of researchers, is unlikely to be
materially affected. Finally, the niche map (Figure~\ref{fig:niche}) uses t-SNE
for visualisation only, and t-SNE neighbourhoods can be layout- and
seed-dependent; the substantive niches are read from the characteristic terms of
each cluster rather than from inter-point distances in the 2-D plot, so they do
not hinge on the projection's stability. Term selection is automatic given the
stop-word list, but that list is itself a manual artefact: extending it beyond
standard function words is what removes person names and filler vocabulary, and
a few methodological or presentational tokens still survive it (Section~S20). The university--topic fingerprints (Figure~\ref{fig:unitopic}) are
likewise descriptive: the shared lean of the most unique researchers towards
computational and machine-learning topics is measured against a clinical-heavy
field baseline and may partly reflect abstract-writing conventions and the
interdisciplinary author profiles that a field-restricted query assigns, rather
than a purely neuroscientific specialisation. These caveats concern the
institutional ranking and presentation; the robust claims (that every university
exceeds the standard, and the researcher-level niche structure) are unaffected.

\subsection{Future work}
Averaging a researcher's paper embeddings collapses them to one point and
discards their topical spread, so interdisciplinary researchers are poorly
represented. A natural next step is to treat each researcher as a distribution
over their papers, which unbalanced optimal transport can compare directly,
letting \rui{} address interdisciplinarity head-on. Because it needs no citations
and reads research content directly, \rui{} may also extend to fields poorly
served by citation metrics, such as the humanities, once adequate text coverage
is secured and non-English and book-length outputs are handled; our replication
in economics, a social science, is a first step. Finally, a disruption-style
index based on forward citations \citep{funk2017,wu2019} would complement the
reference-based check but requires citing--cited network data beyond our current
access, and benchmarking against interdisciplinarity and funding/award outcomes
\citep{stirling2007} is likewise left to future work.

\section{Conclusion}
We introduced \rui{}, an unbalanced-optimal-transport index of research
uniqueness relative to a same-field standard, with an entropy-based
configuration procedure. In neuroscience (2021--2025) all ten studied
universities are substantially more unique than the standard, a result that
replicates in computer science and economics, and each carries a recognisable
research fingerprint at the level of its most unique researchers. We also delimit
the idea: collapsed to one number per institution, the index largely tracks topic
concentration and is sensitive to how an institution's boundary is drawn, so its
ranking should be read with care and its main use is science mapping and
portfolio diagnosis rather than a league table. At the level of individual
researchers, however, \rui{} is not reducible to concentration: it tracks how far
a researcher sits from the field and stays uncorrelated with citation impact. Measuring how much an
institution publishes, and how often it is cited, tells us little about how
different its science is, and that difference, read at the level of
researchers and research areas, is worth measuring. Extending the framework
over time, and to further fields beyond the two studied here, is a natural next
step.

\section*{Data and code availability}
All records were retrieved from Scopus (Elsevier) between June and July 2026.
Under the Scopus data licence, the raw records (abstracts and bibliographic
metadata) cannot be redistributed; they are available to Scopus subscribers, and
Section~S1 lists the affiliation identifiers, subject area, and date range needed
to reproduce every query. All analysis code, together with the derived,
non-restricted data that do not reproduce Scopus content---per-researcher
uniqueness scores, the university-level \rui{} and $\rui_{75}$ values, the fitted
researcher embeddings, and the characteristic-term and topic tables---will be
deposited in a public repository (Zenodo) with a citable DOI on acceptance.

\section*{Author contributions (CRediT)}
\noindent
\textbf{Yu Ota:} Conceptualization, Data curation, Formal analysis,
Investigation, Methodology, Software, Validation, Visualization,
Writing -- original draft.
\textbf{Amane Koizumi:} Conceptualization, Methodology, Project
administration, Supervision, Writing -- review \& editing.
\textbf{Takayuki Ogi:} Conceptualization, Methodology, Supervision,
Writing -- review \& editing.
\textbf{Yo Nakawake:} Conceptualization, Methodology, Supervision,
Writing -- review \& editing.
\textbf{Daichi Mochihashi:} Conceptualization, Methodology, Supervision,
Writing -- review \& editing.

\section*{Funding}
This research was supported by JSPS Kakenhi Grant No.\,26K21781.

\section*{Competing interests}
The authors declare no competing interests.

\section*{Acknowledgments}
This study used bibliographic metadata from Scopus (Elsevier). The analysis
relied on open-source software, including the Python Optimal Transport (POT)
library, NumPy, SciPy, and scikit-learn.

\section*{ORCID}
\noindent
Yu Ota: \url{https://orcid.org/0009-0009-8311-0574} \\
Amane Koizumi: \url{https://orcid.org/0000-0003-1177-0185} \\
Takayuki Ogi: \url{https://orcid.org/0000-0002-5646-4096} \\
Yo Nakawake: \url{https://orcid.org/0000-0002-4911-0740} \\
Daichi Mochihashi: \url{https://orcid.org/0000-0003-0344-5382}


\end{document}